\documentclass[12pt]{iopart}
\usepackage [dvips]{graphicx}
\usepackage{iopams}
\usepackage{epsfig}
\usepackage{verbatim}

\usepackage{color}

\addtocounter{secnumdepth}{1}
\font\capfont=cmbx12 at 50 pt 
\newbox\capbox \newcount\capl \def\a{A}
\def\docappar{\medbreak\noindent\setbox\capbox\hbox{%
\capfont\a\hskip0.15em}\hangindent=\wd\capbox%
\capl=\ht\capbox\divide\capl by\baselineskip\advance\capl by1%
\hangafter=-\capl%
\hbox{\vbox to8pt{\hbox to0pt{\hss\box\capbox}\vss}}}
\def\cappar{\afterassignment\docappar\noexpand\let\a }

\begin{document}

\newcommand{\ee}{{\rm e}}
\newcommand{\dd}{{\rm d}}
\newcommand{\p}{\partial}
\newcommand{\calT}{{\cal T}}
\newcommand{\bex}{\boldsymbol{e}_x}
\newcommand{\bey}{\boldsymbol{e}_y}
\newcommand{\bq}{\mathbf{q}}
\newcommand{\brr}{\mathbf{r}}
\newcommand{\bv}{\mathbf{v}}

\newcommand{\symbe}{{\cal E}}
\newcommand{\symbn}{{\cal N}}
\newcommand{\shadow}{shadow}
\newcommand{\DD}{{D}}
\newcommand{\KK}{{K}}
\newcommand{\DW}{{\DD_W}}
\newcommand{\DB}{{\DD_B}}
\newcommand{\KT}{{\KK_\tau}}
\newcommand{\KTp}{{\KK_{\tau'}}}
\newcommand{\rhoap}{\rho^{\mathrm{ap}}}
\newcommand{\rhofs}{\rho^{\mathrm{fs}}}
\newcommand{\vap}{v^{\mathrm{ap}}}
\newcommand{\vfs}{v^{\mathrm{fs}}}
\newcommand{\KKK}{{\cal P}}
\newcommand{\Kap}{\KKK^{\mathrm{ap}}}
\newcommand{\Kfs}{\KKK^{\mathrm{fs}}}
\newcommand{\Qap}{Q^{\mathrm{ap}}}
\newcommand{\Qfs}{Q^{\mathrm{fs}}}
\newcommand{\PQ}{{\cal Q}}
\newcommand{\rbarap}{\rbar^{\mathrm{ap}}}
\newcommand{\rbarfs}{\rbar^{\mathrm{fs}}}
\newcommand{\Pap}{P^{\mathrm{ap}}}
\newcommand{\Pfs}{P^{\mathrm{fs}}}

\newcommand{\rhobar}{\overline{\rho}}
\newcommand{\rhop }{\rho^\symbe}
\newcommand{\rhom }{\rho^\symbn}
\newcommand{\rhopm}{\rho^{\symbe/\symbn}}
\newcommand{\rhomp}{\rho^{\symbn/\symbe}}
\newcommand{\etabar}{\overline{\eta}}
\newcommand{\etap}{\eta^\symbe}
\newcommand{\etam}{\eta^\symbn}
\newcommand{\etapm}{\eta^{\symbe/\symbn}}
\newcommand{\drhop }{\delta\rho^\symbe}
\newcommand{\drhom }{\delta\rho^\symbn}
\newcommand{\drhopm}{\delta\rho^{\symbe/\symbn}}
\newcommand{\drhomp}{\delta\rho^{\symbn/\symbe}}
\newcommand{\hatrhop}{\hat{\rho}^\symbe}
\newcommand{\hatrhom}{\hat{\rho}^\symbn}
\newcommand{\hatrhopm}{\hat{\rho}^{\symbe/\symbn}}
\newcommand{\hatdrhop}{\delta\hat{\rho}^\symbe}
\newcommand{\hatdrhom}{\delta\hat{\rho}^\symbn}
\newcommand{\hatdrhopm}{\delta\hat{\rho}^{\symbe/\symbn}}
\newcommand{\rbar}{r}
\newcommand{\tilderhop}{\tilde{\rho}^\symbe}
\newcommand{\tilderhom}{\tilde{\rho}^\symbn}
\newcommand{\tilderhopm}{\tilde{\rho}^{\symbe/\symbn}}
\newcommand{\vp}{v^\symbe}
\newcommand{\vm}{v^\symbn}
\newcommand{\vpm}{v^{\symbe/\symbn}}

\newcommand{\ttau}{\tilde{\tau}}
\newcommand{\tk}{\tilde{k}}

\newcommand{\la}{\langle}
\newcommand{\ra}{\rangle}
\newcommand{\beq}{\begin{equation}}
\newcommand{\eeq}{\end{equation}}
\newcommand{\bea}{\begin{eqnarray}}
\newcommand{\eea}{\end{eqnarray}}
\def\lsim{\:\raisebox{-0.5ex}{$\stackrel{\textstyle<}{\sim}$}\:}
\def\gsim{\:\raisebox{-0.5ex}{$\stackrel{\textstyle>}{\sim}$}\:}

\title{Wake-mediated interaction between driven particles crossing a
perpendicular flow}

\author{{J. Cividini and C. Appert-Rolland}\\[5mm]
{\small Laboratoire de Physique Th\'eorique,
Universit\'e Paris-Sud; CNRS UMR 8627,}\\
{\small b\^atiment 210,
91405 Orsay Cedex, France}\\}

\date{\today}

\begin{abstract}
Diagonal or chevron patterns are known to
spontaneously emerge at the intersection of two
perpendicular flows of self-propelled particles,
e.g. pedestrians.
The instability responsible for this pattern formation has
been studied in previous work in the context of a
mean-field approach. Here we investigate the
microscopic mechanism yielding to this
pattern. We present a lattice model study of the
wake created by a particle crossing a perpendicular flow and
show how this wake can localize other particles
traveling in the same direction as a result of an effective interaction mediated by the perpendicular flow.
The use of a semi-deterministic model allows us to characterize analytically the effective interaction between two particles.
\end{abstract}


\maketitle

\section{Introduction}

\cappar
Spontaneous formation of patterns from simple
interaction rules between individual agents (molecules,
particles, humans, etc) has raised much interest in the
statistical physics community. In particular, it is
well-known that perpendicular intersecting flows may
give rise to diagonal patterns. This was observed for
example at the crossing of two perpendicular corridors,
when a unidirectional pedestrian flow is coming from
each corridor, both
experimentally~\cite{hoogendoorn_d2003} and in
simulations~\cite{hoogendoorn_b2003,yamamoto_o2011}.
Another example is the Biham-Middleton-Levine (BML) model~\cite{biham_m_l1992}, that
models urban road traffic in Manhattan-like cities,
and exhibits similar patterns in the free flow phase.
However, it is only recently that the instability
responsible for this diagonal pattern formation was
studied more
systematically~\cite{cividini_a_h2013,cividini_h_a2013}, for a lattice model with two species of particles (or pedestrians), hopping eastward ($\symbe$) or northward ($\symbn$), on a square lattice, and interacting through an exclusion principle.
It was discovered that in certain geometries the
pattern is not strictly diagonal but has the shape of
chevrons. The system was analyzed in terms of mean-field equations believed to correctly
represent the microscopic particle system.

In this paper we focus on understanding how the diagonal pattern emerges from the interactions between particles at the microscopic scale. We shall characterize the effective interaction between two $\symbe$ particles crossing a flow of $\symbn$ particles. It is the perturbation of the density field of $\symbn$ particles by the $\symbe$ particles that mediates the interaction. This kind of environment-mediated interactions have been extensively studied in the context of equilibrium soft matter \cite{likos2001} and more recently in out-of-equilibrium systems \cite{dzubiella_l_l2003,khair_b2007}, with a focus on the prediction of effective forces. In our case, the interaction between particles is not modeled in terms of forces (indeed, the notion of force is not appropriate for, \textit{e.g.} pedestrians) but rather derives from the dynamical rules. A similar question was addressed in \cite{mejia_o2011}, for another lattice model. We shall discuss further in the conclusion how our approach differs from the previous ones and complements them.

We had found in~\cite{cividini_a_h2013,cividini_h_a2013} that pattern formation is generic for updates with low enough noise on the velocity of the particles. Here we consider two of them, frozen shuffle update and alternating parallel update. These updates are deterministic enough to allow for a quantitative description of the effective interaction, not only in terms of ensemble averaged quantities as in \cite{mejia_o2011}, but also at the level of specific realizations of the dynamics.

Our paper is organized as follows. Section~\ref{section:model} gives the definition
of the model and summarizes previous results. The ensemble averaged wake of a single $\symbe$ particle in a uniform background of $\symbn$ particles (i.e. the perturbation
that it induces in the density
of $\symbn$ particles) is computed analytically
in section~\ref{section:ensavg}, showing good agreement
with numerical measurements. Then, in section~\ref{section:micro},
instead of considering an ensemble average,
we shall consider the wake obtained in a given
realization of the model history and characterize
its structure.
This level of description will allow us to obtain
the central result of this paper in section~\ref{section:interaction},
namely the existence and the properties of an effective
interaction between two $\symbe$ particles,
mediated by the perturbation of the density of $\symbn$
particles. Section~\ref{section:ensavg} on the one hand and sections~\ref{section:micro}-\ref{section:interaction} on the other hand can be read independently. Although the calculations of sections~\ref{section:ensavg}, \ref{section:micro} and \ref{section:interaction} are carried out in the case of frozen shuffle update, most of the results are expected to hold for a larger class of update schemes. Section~\ref{section:ap} shows how some of these results can be readily applied to alternating parallel update.
 In section~\ref{section:ccl} discussion and conclusion are presented.

\section{The model}
\label{section:model}

\subsection{Geometry}

The intersection of two perpendicular flows has been modeled using a cellular automaton model. The basic brick of this model is the Totally Asymmetric Simple Exclusion Process (TASEP), a cellular automaton which consists in a directed unidimensional sequence of sites each of which is either empty or occupied by one particle (for review see \cite{chou_m_z2011, derrida1998a,rajewsky1998}). The only enabled motion of a particle is a hop towards the site directly following its current position, happening with probability $p$. New particles can randomly enter the system on the first site with probability $\alpha$ if it is empty and exit it from the last site with probability $\beta$.

Here we will only consider the deterministic version of the bulk dynamics, {\it i.e.} a particle will hop with probability $p = 1$ if its target site is empty. This gives a smoother motion which is expected to be closer to pedestrian transport. We want to avoid any effect from the exit boundaries as well. Particles are therefore allowed to leave with probability $\beta = 1$ if they stand on an exit site. 

\begin{figure}
\begin{center}
\scalebox{.50}
{\includegraphics{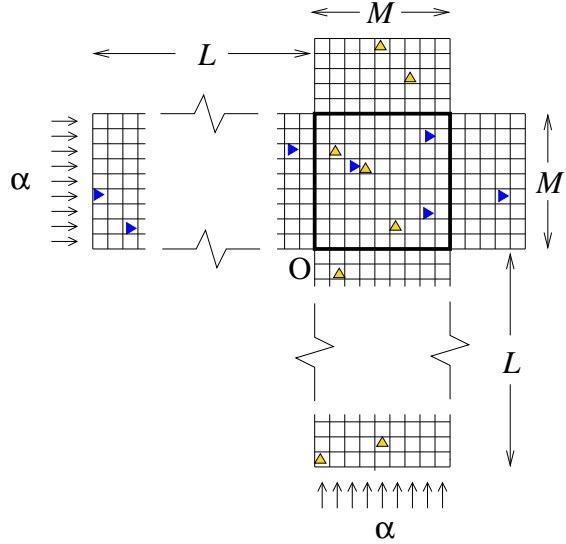}} 
\end{center}
\caption{\small Intersection of two streets of width $M$. The $\symbe$ particles (blue triangles pointing right) move eastward and the $\symbn$ particles (orange triangles pointing up) northward. The parameter $\alpha$ determines the particle injection rate. The `intersection square' is denoted by the heavy solid line.} 
\label{fig_intersectinglanes}
\end{figure}

We call {\it street} an ensemble of several parallel TASEPs. We shall focus on the intersection formed by two perpendicular streets directed toward the east and the north as shown in Fig.\,\ref{fig_intersectinglanes}. A site of the intersection will be denoted by the coordinates $(x,y)$. Each site can be either empty, or occupied by an eastbound ($\symbe$) particle or by a northbound ($\symbn$) particle. For simplicity we consider only the case where there are $M$ TASEPs directed to the north and $M$ to the east, so that the intersection is a square of side $M$. 

We symbolically take the limit $M \rightarrow \infty$, that is to say that in our analytical appoach we will never consider a system with explicit boundaries. However the injection algorithm will still determine the form of the correlations between particles in the bulk.

\subsection{Update scheme}

The properties of the model also depend on the order in which the particles are updated \cite{rajewsky1998}. In most of this paper, we shall use the \textit{frozen shuffle update} \cite{appert-rolland_c_h2011a, appert-rolland_c_h2011b}. It requires to determine a random but fixed order in which the particles will be updated (attempt to hop) at each time step. This is done by giving each particle a {\it phase} $\tau \in [0;1)$, which does not change from one time step to another. Particles are then updated in the order of increasing phase. This is equivalent to updating a particle of phase $\tau$ at all times $t + \tau$, where $t$ is an integer and time is understood as continuous. 

Particles are injected at the each entrance site with constant rate $a$, provided the site is empty. The injection procedure also inserts the new particles in the updating sequence~\cite{appert-rolland_c_h2011b}.
We define the parameter $\alpha$ as the \textit{probability} that a particle enters the system during a time step on a given entrance site when it is empty, so that we have $\alpha = 1 - \ee^{-a}$. The entrance rate determines the density in the free flow phase $\rhofs = \frac{a}{1+a}$~\cite{appert-rolland_c_h2011b}, where the superscript '$\mathrm{fs}$' refers to \textit{frozen shuffle}.

Note that, with our deterministic choices $p=1$ and $\beta=1$, and for the frozen shuffle update, the $\symbn$ particles move with velocity $1$ if there are no $\symbe$ particles in the intersection, \textit{i.e.} no blocking should be seen. This property will be important in the following.

In section~\ref{section:ap} we shall consider another update scheme, the \textit{alternating parallel update}, which we do not detail here. It is however worth noting that the phenomena described in subsection~\ref{subsection:results} are also observed using alternating parallel update.

We also claim that most of the calculations done in the following for a particular update scheme are easily generalizable to other updating schemes, provided the free flow phase is deterministic (moving with velocity $1$) and there is exactly one time unit between two updates of a given particle. But for clarity we shall concentrate from now on, except otherwise stated, on the case of the frozen shuffle update. 

\subsection{Known results}
\label{subsection:results}

Direct Monte-Carlo simulations show that there is a jamming transition at high density\,\cite{hilhorst_a2012}, which is outside the scope of this work. We rather focus on the low-density ($\alpha \lesssim 0.1$) bulk properties of the system, in particular pattern formation. A detailed report of the numerical results can be found in ~\cite{cividini_h_a2013}.  Here we only summarize the results necessary to the understanding of this paper.

First, the model has been simulated on a torus. Particles with random phases are then dropped on the intersection square at initial time. For densities under the jamming threshold, the particles are observed to self-organize into stripes of alternating types directed along the $(1,-1)$ vector. Note that the symmetry with respect to the $(1,1)$ direction on each site of the system imposes the angle of the stripes to be exactly $45^\circ$ (with respect to the vertical).

\begin{figure}
\begin{center}
\scalebox{0.25}
{\includegraphics{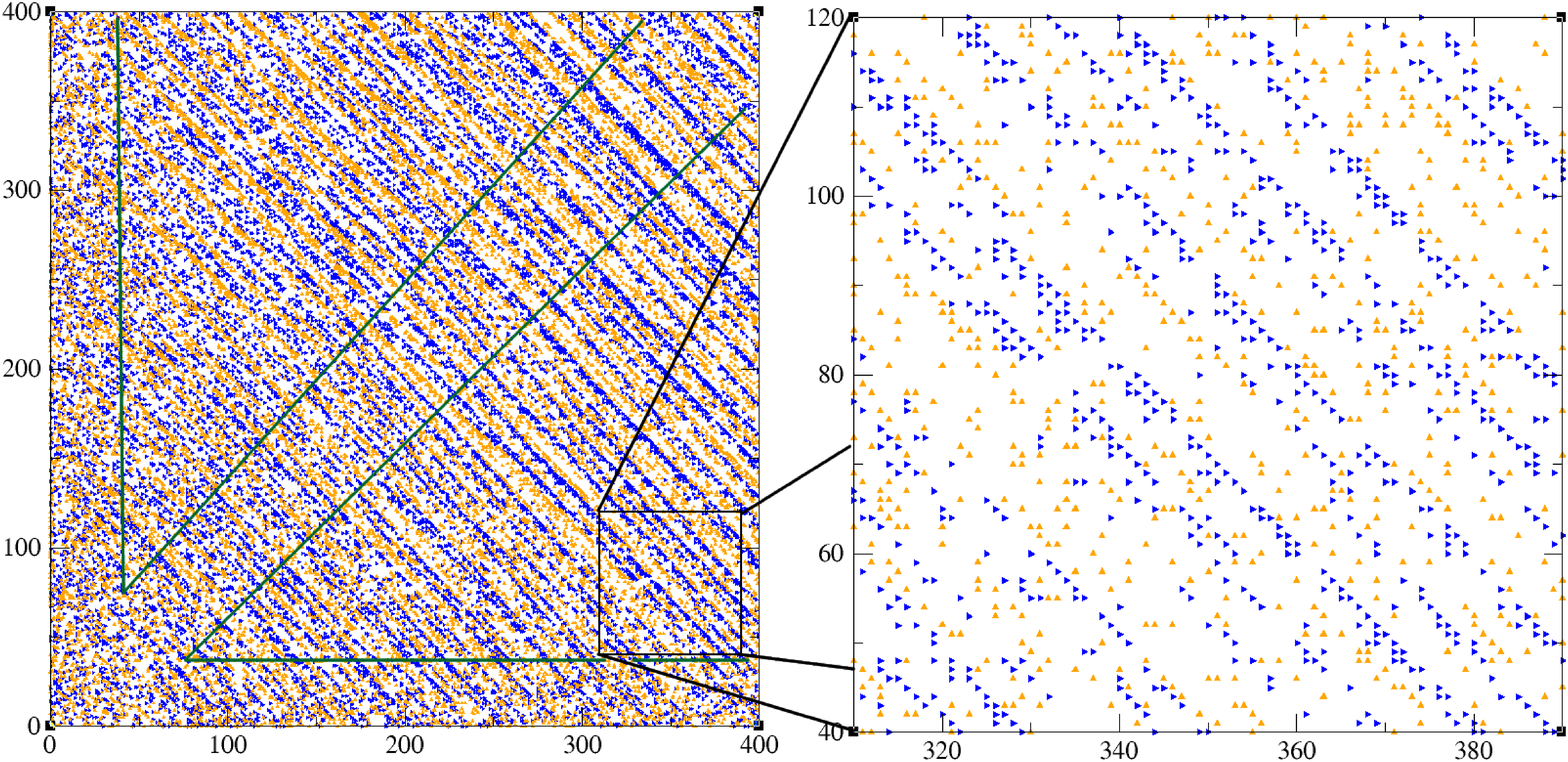}} 
\end{center}
\caption{\small Observed chevron pattern in a simulated open intersection square with $M=400$, $\alpha = 0.105$ and frozen shuffle update. The blue $\symbe$ particles hop towards east and the orange $\symbn$ particle hop towards north. After a penetration length at the boundaries, the particles are observed to self-organize into alternating stripes. These stripes are however slightly tilted with respect to $45^\circ$ depending on the position in the intersection square. In the lower triangle (delimited by green lines), the angle of the stripes with respect to the vertical is somewhat smaller than $45^\circ$, with a deviation of the order of a degree.  The tilt can be linked to an asymmetry in the organization of the types of particles visible in the zoom on the left, \textit{i.e.} in the lower triangle $\symbe$ particles are compacted in alignments whereas $\symbn$ particles are more randomly positioned. 
} 
\label{fig_chevron}
\end{figure}

The more complex case of open boundaries is now considered, schematized in Fig.\,\ref{fig_intersectinglanes}. Stripes are again observed, far enough from the entrance boundaries. However, they now have the shape of \textit{chevrons}. More precisely, the intersection square can roughly be divided in two triangles: above (below) the main diagonal, the stripes are observed to be tilted with respect to $45^\circ$ by a constant amount $\pm \Delta\theta_0$ of the order of a degree and growing with the entrance rate $\alpha$ (see Fig.\,\ref{fig_chevron}). This effect can be qualitatively understood as resulting from the asymmetry in the organization of the two types of particles.

In \cite{cividini_a_h2013,cividini_h_a2013}, a mean-field approach showed how the diagonal pattern can be explained from a linear stability analysis while the chevron effect was shown to be a nonlinear effect. As a complement to this macroscopic approach, we want now to understand through which microscopic mechanism the chevron structure can emerge in the stochastic particle model.

In this paper we describe and quantify how an effective interaction between two $\symbe$ particles can be mediated by the perpendicular flow of $\symbn$ particles. As a first step, we compute the perturbation of the $\symbn$ density field created by a single $\symbe$ particle.

\section{Ensemble averaged wake of a particle}
\label{section:ensavg}

\subsection{Definition of the wake}

An isolated $\symbe$ particle propagating in a flow of randomly incoming $\symbn$ particles will create a perturbation in the average $\symbn$ density. The ensemble-averaged density pattern seen in the frame moving with the $\symbe$ particle will be called its {\it wake}. 
In this section we compute the shape of the wake for the frozen shuffle update, although the method is intended to be more general.

The system is taken to be infinite so it is translationally invariant. Throughout this section, the moment at which the $\symbe$ particle has hopped to a site taken to be $(0,0)$ will be chosen as the time origin $t=0$ without loss of generality. The stationary state was established in the negative times. We want to compute the stationary ensemble-averaged density of $\symbn$ particles on any site $(k,l)$ at time $0$. We call it $\rhom_{kl}$. The dependence of $\rhom_{kl}$ on the incoming $\symbn$ particle density $\rho$ will be implicit in the following.

Before performing the ensemble average,
we need to compute the density field of $\symbn$ particles
for a given realization of the dynamics.
This will be the aim of the next subsection.

\subsection{Density field for a given realization
of the $\symbe$ particle dynamics}
\label{sec:wake_not_averaged}

One first remark is that, for a given realization,
the past dynamics of the $\symbe$ particle
is fully defined by the number of time steps the particle
spent on each of the already visited columns $p=-1,-2,\ldots$,
which we call $\{n_p\}$.
Several realizations of the $\symbn$ particles dynamics
may yield the same values for the $\{ n_p \}$'s.
We average over all these realizations.
We are thus considering in this section a given
realization
of the $\symbe$ particle dynamics, averaged over all
the compatible realizations of the $\symbn$ particles' dynamics.

A second remark is that, when the $\symbe$ particle leaves
a column after perturbing the density in it, the
density pattern in that column simply evolves
independently of the subsequent hops of the $\symbe$ particle.
Thus it will be convenient to decompose the density
field into density profiles in each column that
will evolve independently in time.

Let us consider a column that particle $\symbe$ visited in
the past, corresponding to the sites $(k,l)$ with
$k<0$ fixed, and $l$ ranging from $-\infty$ to $+\infty$.
The $\symbe$ particle left column $k$ at time $-\sigma_k \equiv - \sum^{-1}_{k^\prime = k+1} n_{k^\prime} \le 0$.
We shall show now that the density profile in column $k$
and at time $-\sigma_k+s$ ($s\ge0$)
depends only on three parameters:
$n_k$, $n_{k-1}$, and $s$
(the variable $s$ counts how many time steps were elapsed
since particle $\symbe$ left column $k$).
As a consequence, we shall denote this density profile
by $R^{n_{k-1} n_k}_l(s)$.
We insist that this quantity is not explicitly dependent on $k$.

Let us first determine $R^{n_{k-1} n_k}_l(0)$.
Particle $\symbe$ was blocked $n_{k-1}-1$ time steps
before entering column $k$, due to the presence
of $\symbn$ particles on column $k$.
As, before the arrival of the $\symbe$ particle,
the $\symbn$ particles are in free flow (i.e. moving with
velocity $1$), this indicates that there were $n_{k-1}-1$
adjacent $\symbn$ particles on column $k$.
This platoon was not altered by the $\symbe$ particle and continues to move
with velocity $1$ at all subsequent time steps.

\begin{figure}
\begin{center}
\scalebox{0.7}
{\includegraphics{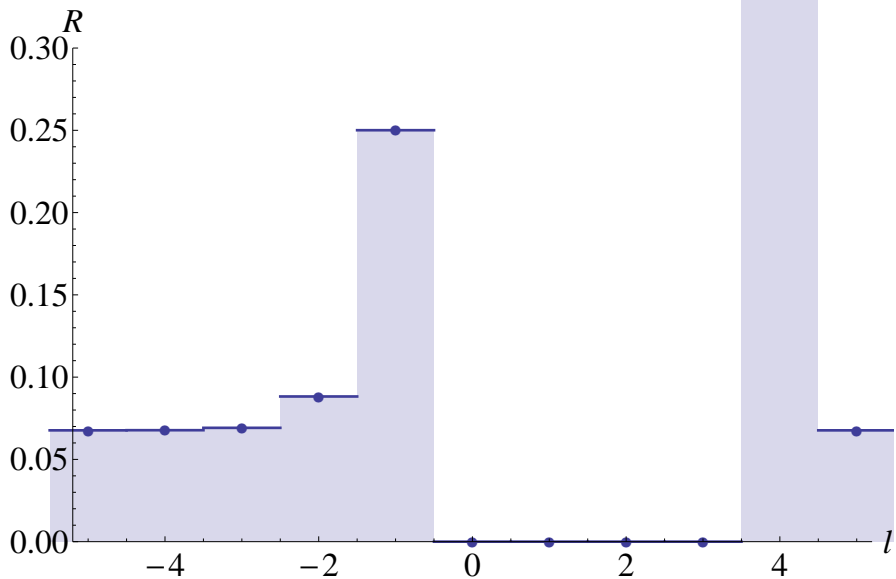}} 
\end{center}
\caption{\small Function $R^{23}_l(0)$ for an incoming $\symbn$ particle density $\rho = 0.067$.
For $l\ge5$, the density is equal to $\rho$ and has not
been perturbed by the $\symbe$ particle. 
Site $4$ is surely occupied, i.e.  $R^{23}_4(0) = 1$, because the $\symbe$ particle was blocked once before
entering the column. There is an empty region of size $3 + 1 =
4$ for $0 \leq l < 4$ corresponding to $\symbn$ particles being
blocked by the $\symbe$ one. The blocked $\symbn$
particles then accumulate in the region $l < 0$, where the
density is larger than $\rho$.
However, the density quickly decreases to its asymptotic value
$\rho$ as $l$ goes towards negative values.
} 
\label{fig_R0}
\end{figure}

Particle $\symbe$ hopped on column $k$ just behind this
platoon, and then blocked the column for $n_k$ time steps.
The next incoming $\symbn$ particles were forced to queue up.
This created an empty zone of size $n_k+1$ for $l\ge0$,
and a denser zone for $l<0$.

As a summary, the density profile $R^{n_{k-1} n_k}_l(0)$
just after particle $\symbe$ left column $k$ reads:
\begin{equation}
R^{n_{k-1} n_k}_l(0) = \left\{
\begin{array}{cl}
\rho & \mbox{for } l \ge n_{k-1}+n_k  \\
1    & \mbox{for } n_k < l < n_{k-1}+n_k  \\
0    & \mbox{for } 0 \le l \le n_k  \\
\ge \rho  & \mbox{for } -\infty \le l < 0  \\
\end{array}
\right.
\label{eq:R}
\end{equation}

In principle, the $\symbe$ particle can block an arbitrary large number of $\symbn$ particles. However, in the low density regime, a $\symbn$ particle is added to the queue with a probability $\sim \rho$. The density profile thus decreases very rapidly to its asymptotic value $\rho$ for $l < 0$.
An exemple of such a density profile is given in
Fig.\,\ref{fig_R0}.
The explicit calculation of the profile for $l<0$ is described
in~\ref{section:queue}.

Once the $\symbe$ particle has left column $k$, the particles
queuing at $l < 0$, if any, may have undergone a short transient until a new free flow configuration
was reached.
Here we study low densities and thus we have
neglected the possible modification of the density profile in
the transient.
After this transient the density profile in column $k$ moves
upwards unchanged,
which enables us to compute the $R^{n_{k-1} n_k}_l(s)$ for
all $s\ge0$.

\subsection{Ensemble average of the wake}

For $k\le0$, one can compute the wake $\rho^\symbn_{kl}$
by averaging $R^{n_{k-1} n_k}_l(s)$
on the past dynamics of the $\symbe$ particle.
The probability $\rbar_{n_p}$
associated with each possible value of $n_p$ is proved in \ref{section:rbar} to be equal to
\beq
  \rbarfs_n = \left\{
  \begin{array}{l l}
    1-\rhofs & \quad n = 1 \\
    \frac{\rhofs}{a} \ee^{-a} \displaystyle \sum_{l=n-1}^\infty \frac{a^l}{l!} & \quad n=2,3,\ldots \\
  \end{array}. \right.
\label{eq:rbarfs}
\eeq

For $k > 0$, i.e. in the columns that have not yet been reached
by the $\symbe$ particle, $\rho^\symbn_{kl}$ has to be equal to
the $\symbn$ particle density $\rho$. We thus have
\beq
\rho_{kl}^\symbn = \left\{
  \begin{array}{l l}
     \displaystyle \sum_{n_{k-1} = 1}^{\infty} \sum_{n_{k} = 1}^{\infty} \cdots \sum_{n_{-1} = 1}^{\infty} \rbar_{n_{k-1}} \rbar_{n_{k}} \cdots \rbar_{n_{-1}} R^{n_{k-1}n_k}_l (\sum_{k'=k+1}^{0} n_{k'}) & \quad k \leq 0 \\
   &  \\
    \rho& \quad k > 0 \\
  \end{array} \right.
\label{eq:rhokl}
\eeq
Note that $\rhom_{kl}$ is computed when the $\symbe$ particle just entered the $k=0$ column, which corresponds to taking $n_0 = 0$.

\begin{figure}
\begin{center}
\scalebox{0.3}
{\includegraphics{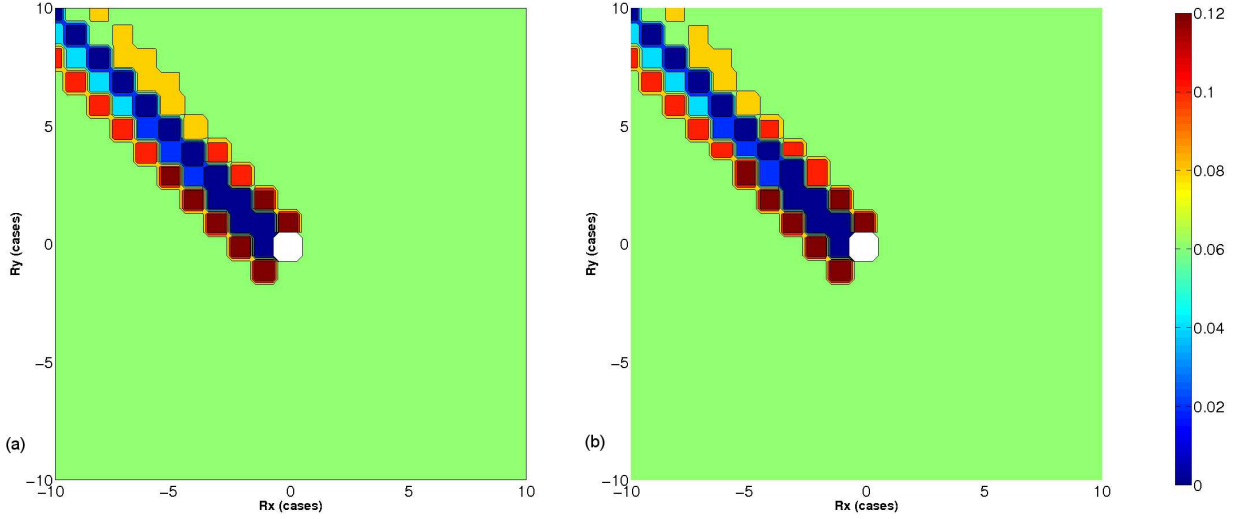}} 
\end{center}
\caption{\small Colorplot of the density perturbation (wake) created by a single $\symbe$ particle that just arrived on the $(0,0)$ site (white spot) for $\rho = 0.068$ using frozen shuffle update. (a)~Theoretical wake as calculated in Eq.(\ref{eq:rhokl}). (b)~Numerical measurement of the density field immediately after each update of the $\symbe$ particle, averaged over $5.10^6$ timesteps.
} 
\label{fig_rhoklthexp}
\end{figure}

A comparison between the theoretical and the numerically measured shape of the wake $\rho_{kl}^\symbn$ is shown in Fig.\,\ref{fig_rhoklthexp}.
They are in excellent agreement, justifying the approximation done in section~\ref{sec:wake_not_averaged} where we neglected the relaxation of the density profile after the passage of the $\symbe$ particle.

\section{Microscopic structure of the wake}
\label{section:micro}

Our aim is now to understand how a second $\symbe$ particle can be localized by the wake of a first $\symbe$ particle. As the second particle will not see the average wake but a specific realization of it, we shall now describe in more details its microscopic structure. In particular, we have seen in section~\ref{sec:wake_not_averaged}
that the passage of an $\symbe$ particle creates an empty zone.
We shall now define an algorithm that allows to track the
empty sites for a given realization of the dynamics of the $\symbe$ and $\symbn$ particles. Our algorithm tracks effectively empty sites only for updates such that all the unblocked $\symbn$ particles
move one step forward at each time step (this is indeed the case for the two updates
considered in this paper). 

\begin{figure}
\begin{center}
\scalebox{0.35}
{\includegraphics{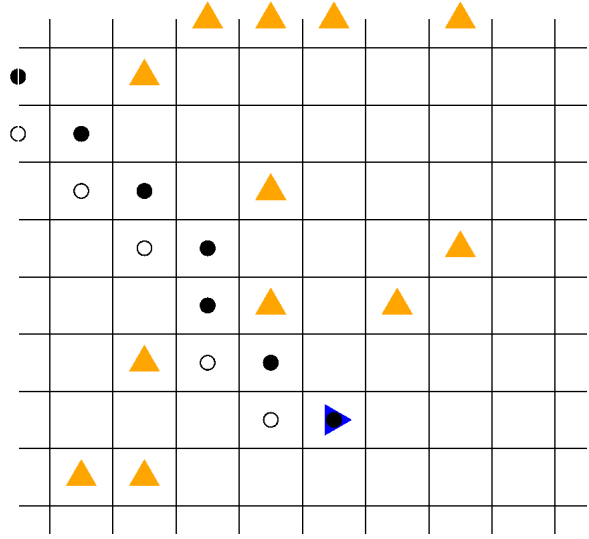}} 
\end{center}
\caption{\small The shadow of a single $\symbe$ particle as defined in section~\ref{section:micro}. The blue right-pointing triangle represents the $\symbe$ particle and the orange up-pointing triangles are the $\symbn$ particles. Dotted sites are denoted by black and white circles. The snapshot is taken directly after the hop of the $\symbe$ particle.
} 
\label{fig_pearls}
\end{figure}

\vspace{2mm}

\subsection{Tracking algorithm}

We define the following algorithm~:

(1) Just before the $\symbe$ particle attempts to move, put a white dot on the site it occupies.

(2) After the move is performed, put a black dot on the site occupied by the $\symbe$ particle which will replace the white dot if the $\symbe$ particle did not move.

(3) During the next time step, let the $\symbn$ particles hop, and 
let all the dots move one site upwards simultaneously.
Start again at step (1).

We call \textit{shadow} the set of the sites occupied by a dot.

All sites in the shadow are empty
for all time steps.
After being created, the shadow is just translated
upwards, and its shape is kept invariant.

\vspace{2mm}

\subsection{Properties of the shadow}

For a single $\symbe$ particle having an infinite history, an infinitely
long line of dots is constructed.
The form of the
shadow is depicted in Fig.\,\ref{fig_pearls}. Note that
according to the algorithm the shadow is well defined at the
instant of the hop of the $\symbe$ particle. In the following,
we shall always consider it just after the hop of the $\symbe$
particle.

Two subsequent black dots are separated by a vertical
line each time the $\symbe$ particle was blocked,
and by a diagonal line each time the $\symbe$ particle moved
forward.
The asymptotic
angle $\theta$ between the shadow and the $y$ axis is
therefore related to the average velocity $v$ of the
$\symbe$ particle, through the relation $\tan \theta = v$.
In the limit of small density, the average velocity behaves like
$1-\rho$.
If $\Delta\theta$
measures the deviation of the angle $\theta$ from $45^\circ$, we
thus have to lowest order in density $ |\Delta\theta| =
\rho/2\,\mathrm{rad} = \rho \times (90/\pi) \mathrm{deg}$. This formula is coherent with the one found in
\cite{cividini_a_h2013} in an ideal case.

For the next section, it will be also useful to notice
that the shadow can only have a width of $1$ or $2$ dots in
each row.
It can be seen as 
a superposition of two types of rows:
from right to left, there are either a
black site followed by a white site or a $\symbn$ particle
followed by a black site.
This $\symbn$ particle is precisely the one that blocked
the $\symbe$ particle in the past.
We call these two types of rows $\DD$ or $\KK$ as defined
in Fig.\,\ref{fig_states}a.

The relative position of two adjacent rows is not arbitrary: below the black dot of a given row, one finds necessarily the leftmost dot (black or white) of the row below.

\section{Wake-mediated interaction with another particle}
\label{section:interaction}

A second $\symbe$ particle will be said to be in the
shadow
of a first $\symbe$ particle if it occupies any of the dotted
sites.
The dynamics of the first particle is entirely encoded in the
shape of the shadow, it therefore suffices to study the
dynamics of the second particle in the shadow to get
informations on the correlations between the $\symbe$
particles.
In particular, we are interested in knowing how long
the second particle will stay in the shadow of the
first one.

\begin{figure}
\begin{center}
\scalebox{0.5}
{\includegraphics{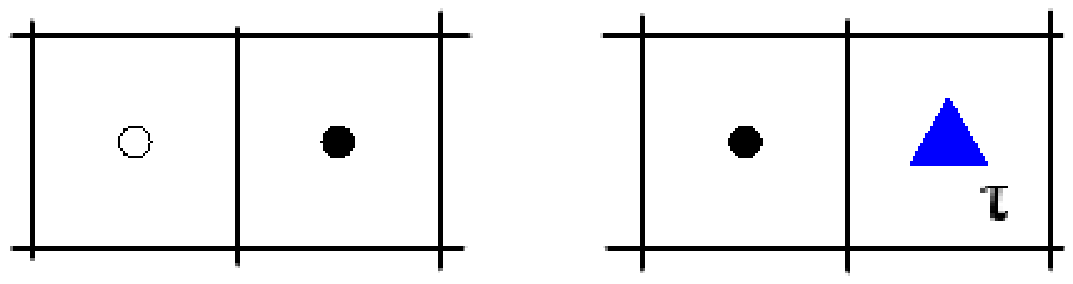}} 

(a)
\hskip 3.5cm
$\DD$ \hskip 2.7cm $\KK$
\hskip 3.5cm
\mbox{      }

\vskip 0.8cm

\scalebox{0.5}
{\includegraphics{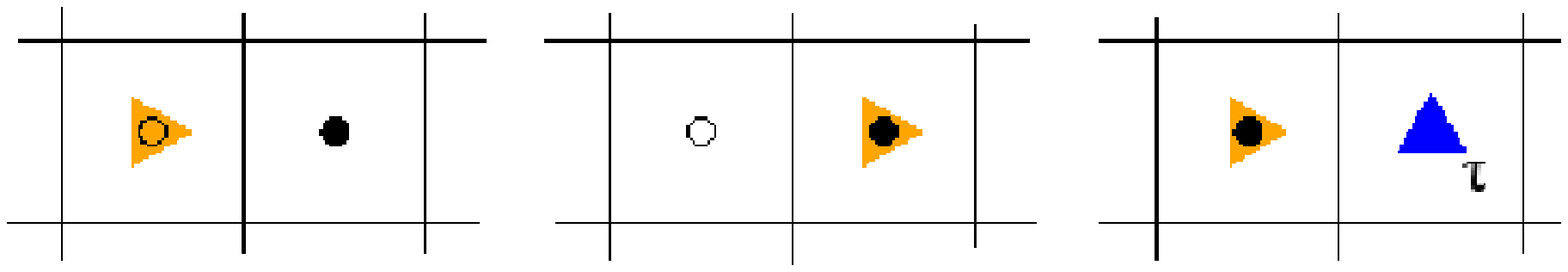}} 

(b)
\hskip 1.5cm
$\DW$ \hskip 2.6cm $\DB$ \hskip 2.6cm $\KT$
\hskip 1.5cm
\mbox{     }
\end{center}
\caption{\small
(a) Possible types of rows.
Type $\DD$ corresponds to having two dots (a white and
a black) in the same row, while type $\KK$ contains
only one black dot, associated with an $\symbn$ particle
on its right.
(b) Possible states of an $\symbe$ particle in the
shadow. If the row is of type $\DD$, the $\symbe$ particle
can be either on a white dot (state $\DW$), or on a
black dot adjacent to a white one ($\DB$).
If the row is of type $\KK$, the $\symbe$ particle
can only be on the black dot.
In the case of the frozen shuffle update, an extra
parameter is the phase $\tau$ of the $\symbn$ particle,
hence a continuum of states $\KK_{\tau \in [0;1)}$.
The algorithm creating and moving the dots is explained in section~\ref{section:micro}.
} 
\label{fig_states}
\end{figure}

We have just seen that the shadow is a superposition
of two types of rows, one with two dots and the other with one
dot and one $\symbn$ particle.
The second $\symbe$ particle can occupy any of these dotted
sites, and as a result can be in three different states,
depicted in Fig.\,\ref{fig_states}b.

We want now to determine whether the second $\symbe$ particle
remains in one of these states or exits the shadow, and how
the transitions are between the
different states.
At each time step, the rows of the shadow are moving upwards with the $\symbn$ particles,
while the second $\symbe$ particle remains on the same row,
and hops forward if the target site is empty.
Thus, in the frame moving upwards with the shadow,
the shadow itself is invariant and the $\symbe$ particle
moves one step in diagonal in the right-bottom direction.
One has to determine whether this step in diagonal is possible,
and whether the $\symbe$ particle will still be in the shadow
at the next time step.
All the subsequent discussions will be in this
shadow-correlated frame.

A first remark is that once the second $\symbe$ particle is in the shadow,
it can never leave the shadow by being left behind.
The second $\symbe$ particle can only
exit the shadow (if it does) by overtaking it.
This is a consequence of the fact that there is always
a dot (black or white) below a black dot.
Indeed, if particle $\symbe$
is on a black dot (states $\DB$ or $\KT$), and if it is
blocked at the next time step, it will still arrive on
a dotted site, and thus
will not exit the shadow from behind.
If it is on a white site (state $\DW$), it will surely
hop forward, and again will not be left behind the shadow.
This is one of the most striking
effect of the shadow on the dynamics of the second particle.

Our aim is to estimate the probability that the second $\symbe$
particle stays in the shadow of the first one after 
$t$ time steps,
averaged over all possible realizations of the flow of
$\symbn$ particles, i.e. of the shadow.
In this section, we shall calculate this probability
in the case of the frozen shuffle update
and denote it $\Pfs_S(t)$.
The phase of the
$\symbe$ particle creating the shadow is taken to be~$0$.
The phase of the second $\symbe$ particle is denoted
as~$\tau_0$.

We have listed in Fig.\,\ref{fig_states}b the possible
states of a particle in a shadow.
As long as the second particle remains in the shadow,
it is necessarily in one of these three states.
Starting from a given initial state, 
the probability to remain in the shadow after $t$ time steps
is thus given by the sum
\begin{equation}
P_S(t) = P_W(t) + P_B(t) + \int_{0}^1 d\tau\; p_\tau(t)
\label{eq:defPS}
\end{equation}
where $P_W(t)$, $P_B(t)$ and $p_\tau(t)d\tau$ stand for the
probabilities of the particle being in state $\DW$, $\DB$ and
$\KK_{\tau' \in [\tau, \tau+d\tau[}$
at time $t$, respectively.
We have to write some rate equations
for these probabilities based on
the microscopic dynamics.

We shall now present an intuitive explanation of the way rate
equations between the possible states of a particle in
the shadow can be written to linear order in
$\rhofs$ (Eqs.\,\ref{eq:ratelowd}). The more general equations valid for all densities
will be derived 
in~\ref{section:rateeq}.

Suppose we have an $\symbe$ particle in the state
$\DW$, {\it i.e.} occupying a white dotted site. The
probability that the row directly under the particle is
a $\DD$ row is $(1-\rhofs)$. In that case the particle
will stay in state $\DW$ at the next time step. The
particle can arrive in state $\KK_{\tau > \tau_0}$ if
the row under it is a $\KK$ row, with $\tau > \tau_0$
(probability $\tau_0 \rhofs$). Similarly, it will
arrive in state $\KK_{\tau < \tau_0}$ with probability
$(1-\tau_0) \rhofs$. This completes the list of
possible arrival states for a particle leaving state
$\DW$.

If the $\symbe$ particle departs from state $\DB$, we
have to specify if the site directly to its right is occupied
or not. It is empty with probability $(1-\rhofs)$, in which
case the particle can arrive either in state $\DB$ if the row under it
is a $\DD$ row (probability $(1-\rhofs)$), or get blocked by a
$\symbn$ particle if the row under it is a $\KK_{\tau <
\tau_0}$ row (probability $\tau_0 \rhofs$), or exit the
shadow if the row under it is a $\KK_{\tau > \tau_0}$ row 
(probability $(1-\tau_0) \rhofs$).
If, in the initial configuration, there is a $\symbn$
particle on the site to the right of the $\symbe$ particle, the
next row is necessarily a $\DD$ row to linear order in
$\rhofs$. The $\symbe$ particle will therefore stay in the
$\DB$ state if it is not blocked by the $\symbn$ particle (probability $\tau_0 \rhofs$)
or arrive in the $\DW$ state if it is blocked (probability $(1-\tau_0)
\rhofs$).

Finally, we notice that an $\symbe$ particle occupying a
state $\KK_{\tau < \tau_0}$ will necessarily arrive in state
$\DB$ at the next time step and that a particle occupying a
state $\KK_{\tau > \tau_0}$ will arrive in $\DW$. We can
therefore write the rate equations to linear order in $\rhofs$:
\beq
\left\{
\begin{array}{l l l l}
\Pfs_W(t+1) & = (1-\rhofs) \Pfs_W(t) + \rhofs (1-\tau_0) \Pfs_B(t)+ \Pfs_{> \tau_0} (t) \\
\Pfs_B(t+1) & = (1-2 \rhofs + \rhofs \tau_0) \Pfs_B(t) + \Pfs_{< \tau_0}(t) \\
\Pfs_{< \tau_0}(t+1) & = \rhofs \tau_0 \Pfs_W(t) + \rho \tau_0 \Pfs_B(t) \\
\Pfs_{> \tau_0}(t+1) & = \rhofs (1-\tau_0) \Pfs_W(t)
\end{array} \right.
\label{eq:ratelowd}
\eeq
where we have defined $\Pfs_{< \tau_0} (t) \equiv \int_{\tau =
0}^{\tau_0} p_\tau (t) d\tau$ and $\Pfs_{> \tau_0} (t) \equiv
\int_{\tau = \tau_0}^{1} p_\tau (t) d\tau$.
Therefore, the probability to stay in the shadow
evolves as
\begin{equation}
\Pfs_S(t+1) = \Pfs_S(t)  -\rhofs (1-\tau_0)  \Pfs_B(t)
\label{eq:pfs_lowd}
\end{equation}
and clearly decreases with time.

\begin{figure}
\begin{center}
\scalebox{0.5}
{\includegraphics{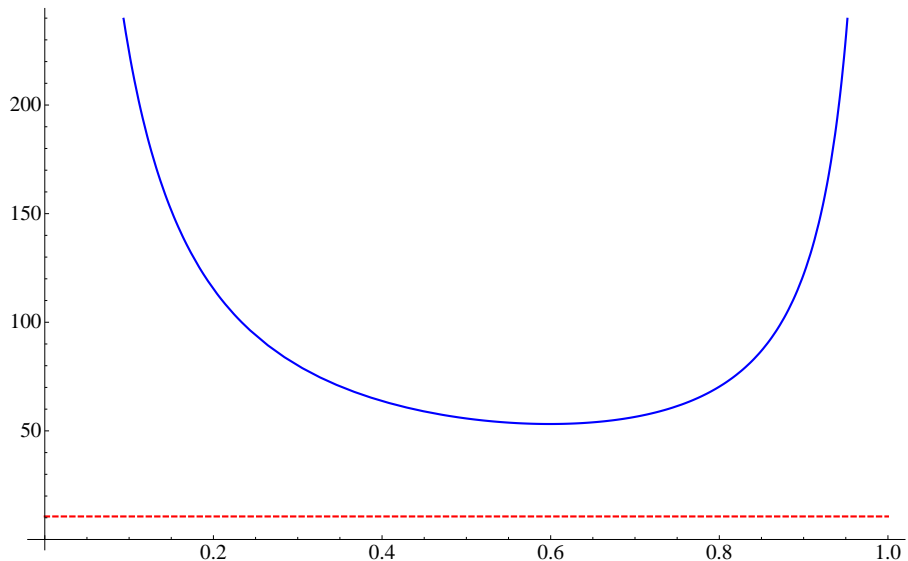}} 
\end{center}
\caption{\small Characteristic escape time of a particle starting from state $\DB$ in the
shadow as a function of its phase $\tau_0$ (blue). The
escape time diverges for $\tau_0=0$ and $1$.
The
escape time in case of uncorrelated particles has been plotted
in red for comparison.
} 
\label{fig_escapetau0}
\end{figure}

\begin{figure}
\begin{center}
\scalebox{0.5}
{\includegraphics{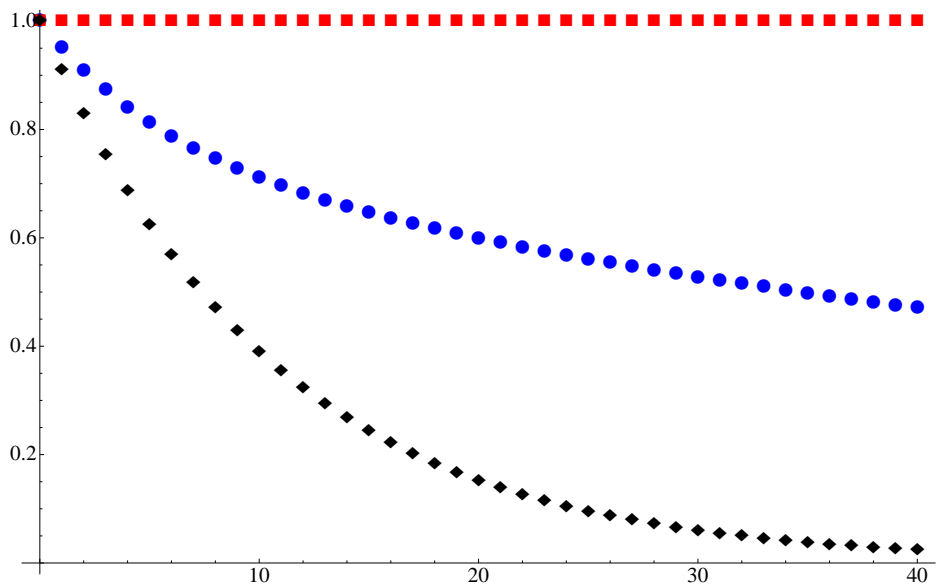}} 
\end{center}
\caption{\small Probability that the second particle stays in
the shadow during the first $t$ timesteps for frozen shuffle
update (blue disks, Eqs.(\ref{eq:ratelowd}-\ref{eq:pfs_lowd})) and alternating parallel
update (red squares, Eqs.(\ref{eq:ratealtpar})). For comparison, the same
quantity has been plotted assuming the two particles are
uncorrelated (black diamonds). The density $\rho$ of $\symbn$ particles
is $0.1$.
}
\label{fig_wakedecay}
\end{figure}

It can be seen from these equations that the
probability $\Pfs_S(t)$ 
depends on the phase
$\tau_0$ between the two $\symbe$ particles.
For one given $\tau_0$ value, one can solve the linear
equations~(\ref{eq:ratelowd}-\ref{eq:pfs_lowd}) by diagonalizing the transfer matrix to obtain the time evolution
of $\Pfs_S(t)$. 
The solution is
given by a linear combination of exponentials\footnote{While the characteristic escape time is independent
of the initial condition, the short-time
evolution of $\Pfs_S(t)$ does depend on it.
In Fig.\,\ref{fig_wakedecay}, we have averaged the
distribution $\Pfs_S(t)$ over all $\tau_0$ values,
while assuming that the initial state was a $\DB$ state.
If we would have considered an initial state $\DW$,
the slope of the curve at $t=0$ would have been horizontal, in
accordance with the fact that the shadow cannot be exited
directly from state $\DW$.}. The longest of the characteristic decay times has been
plotted as a function of $\tau_0$ in Fig.\,\ref{fig_escapetau0}.

It is worth underlining that the distribution $\Pfs_S(t)$
as given in Fig.\,\ref{fig_wakedecay} does {\em not}
depend on the distance between the two $\symbe$ particles, as distance only acts as a delay in the interaction between the $\symbe$ particles.

The probability to stay in the shadow is compared to a reference situation where
the second $\symbe$ particle meets an unperturbed flow of $\symbn$ particles (the shape that the
shadow would have if the flow was perturbed by a first $\symbe$ particle is computed only to define the zone from
which the exit time is measured). In this reference situation, the displacements of both
$\symbe$ particles are uncorrelated.
We find that for any $\tau_0$ value, 
$\Pfs_S(t)$ decays more slowly if the motion of the two
$\symbe$ particles is correlated through the mediation
of the shadow, than for decorrelated motion (see
Fig.\,\ref{fig_escapetau0}).
In particular, with the shadow, the escape time diverges when
$\tau_0$ becomes close to $0$ or $1$.
As a result, when averaged over $\tau_0$ (see
Fig.\,\ref{fig_wakedecay}),
the decay in the uncorrelated case is exponential
whereas the probability to stay in the shadow decays like
$t^{-1}$. Indeed, in the large-time limit the average over
$\tau_0$ is dominated by the diverging escape times.

The fact that the exit time
is larger in the presence of a shadow shows the relative
stability of the two-particle state.
Thus we can consider as a bound state with finite life-time
the set of two $\symbe$ particles, one being in the shadow of
the other.

For random initial positions of the $\symbe$ particles (not
necessarily in the shadow), various
types of collisions can destroy the structure of the shadow,
so that
in the case of the crossing of two perpendicular flows
as in Fig\,\ref{fig_intersectinglanes},
the system will not converge towards the pure mode
that we have just described.
However, this state was observed to be realized locally
in direct simulations~\cite{cividini_a_h2013}.

\section{Alternating parallel update}
\label{section:ap}

We want to illustrate the generality of the calculations and discussions of the previous sections~\ref{section:ensavg},\ref{section:micro} and \ref{section:interaction} by applying them to alternating parallel update. Under {\it alternating parallel update}, a parallel update is performed alternatively on $\symbe$ and $\symbn$ particles at each half time step. More precisely, we define the time scale such that $\symbe$ particles move at integer time steps, while $\symbn$ particles move a half time step later.

The wake of a single $\symbe$ particle has mostly the same structure as for the frozen shuffle update. The $R^{n_{k-1} n_k}_l(s)$ function defined in section~\ref{section:ensavg} still verifies the properties detailed in Eq.\,(\ref{eq:R}), although the exact shape is different for $l < 0$. One must also take into account that the mean density in free flow phase as a function of the entrance probability is now $\rhoap = \frac{\alpha}{1+\alpha}$\cite{rajewsky1998} (the superscript '$\mathrm{ap}$' standing for alternating parallel). Again, $\alpha$ stands for the probability to inject a $\symbn$ particle in an entrance site during a time step at which it is empty. Eq.\,(\ref{eq:rhokl}) is also verified provided one uses the correct expression for the $\rbarap_n$, namely
\beq
  \rbarap_n = \left\{
  \begin{array}{l l l}
    1-\rhoap & \quad n = 1 \\
    \rhoap & \quad n=2 \\
    0 & \quad n > 2 \\
  \end{array}. \right.
\eeq

The tracking algorithm defined in section~\ref{section:micro} is still valid as well. Since all the $\symbn$ particles move at the same time, the continuum of states $\KK_\tau$ becomes a single state $\KK$. The equivalents of equations~(\ref{eq:ratelowd}) can then be obtained by setting $\tau_0 = 1$, \textit{i.e.} all the $\symbe$ particles both move at the same time. The calculations for alternating parallel update being formally obtained as a special case of the frozen shuffle update, all the remarks made in section~\ref{section:interaction} also hold here. We get for the evolution equations
\beq
\left\{
\begin{array}{l l l}
\Pap_W(t+1) & = (1-\rhoap)\Pap_W(t) \\
\Pap_B(t+1) & = (1-\rhoap) \Pap_B(t) + \Pap_{\KK}(t) \\
\Pap_{\KK}(t+1) & = \rhoap \Pap_W(t) + \rhoap \Pap_B(t)
\end{array} \right.,
\label{eq:ratealtpar}
\eeq
which are in fact true for all densities, as shown in~\ref{section:rateeq}.
One can see that in this case,
the second $\symbe$ particle cannot exit
the shadow. Indeed, Eqs.(\ref{eq:ratealtpar}) sum up
to $\Pap_S(t+1) = \Pap_S(t)$, and 
the decay time is infinite.

Another way to phrase it is that it
is impossible for the second $\symbe$ particle to leave a black
site (state $\KK$ or $\DB$) for a non-black one
(state $\DW$ or exit from the shadow).
A consequence is that
if the second particle stands on a black site,
its shadow coincides with the shadow of the first particle.
We can then add more than one particle. In particular, the
state in which all black sites are occupied by an $\symbe$
particle is a stable state consisting in an infinite line of
$\symbe$ particles. We recover a macroscopic mode that
had already been proposed in~\cite{cividini_a_h2013} as an explanation for the chevron effect.

\section{Conclusion and discussion}
\label{section:ccl}

The work presented in this paper was triggered by the
observation of an instability at the crossing of two
perpendicular flows, leading to the formation of stripes
that have the shape of chevrons.
While a mean-field approach was proposed  
in~\cite{cividini_a_h2013} and developped
in~\cite{cividini_h_a2013},
we are interested here by the mechanisms involved
at the microscopic scale.

We demonstrate how interactions between
particles of the same type can arise from the mediation
of the perpendicular flow.
In a first stage, we have studied the wake created by
a single particle moving in such a perpendicular flow.
The averaged wake was predicted analytically, in good
agreement with simulations.
The microscopic structure of a given realization of the
wake was also provided, and allowed to show that a second
particle could be localized in the wake
of a first particle.
The localization time depends on the type of update
that is used.
The angle of the wake is the same as the one of the
long-lived global mode identified
in~\cite{cividini_a_h2013}.

The calculations here were done for the frozen shuffle update. We have shown in the last section how it could be generalized to the alternating parallel update. In fact, calculations can be extended to other update schemes, provided the free flow phase is deterministically shifted forward with velocity $1$ at each time step. 

We have also assumed that the flow of $\symbn$ particles
was homogeneous.
In the full problem of crossing flows, the $\symbn$ particles
themselves will be organized into stripes.
As long as these particles move with unit velocity,
our calculation leading to a localization phenomenon
of one $\symbe$ particle in the wake of another
one can be easily generalized.
However, we have not described here how, once it has left
the wake, the second particle can alter the wake of
the first one and also modify the density of
$\symbn$ particles.
In the complete setting, where a whole flow of
$\symbe$ particles crosses the flow of $\symbn$
particles, multiple collisions will result into
a finite length for the wakes.
This length depends on the
type of update and it would be interesting to estimate it.
The angle of the wake may also be altered by these
collisions. The resolution of the full problem would
require to be able to estimate this new angle - though
the order of magnitude should not be modified compared
to what was found here, as already mentioned
in~\cite{cividini_a_h2013}.

Of special interest is the alternating parallel
update, for which a particle can be localized in the
wake of another for infinite times.
As a result, the stripes of the global pattern
observed in the complete problem are more contrasted than for frozen shuffle update \cite{cividini_h_a2013}.

While the analytical work presented in this paper
was first triggered by the observation of patterns
in pedestrian
crossings, it can also be cast into the more general
research field of effective interactions.
These effective interactions mediated by the environment
were first studied in soft condensed matter physics
for systems at equilibrium (see a review in \cite{likos2001}).
A classical example is the ``depletion attraction''
due to entropic effects, that appear between two large colloidal
particles placed in a dilute bath of smaller colloids~\cite{asakura_o1958}.
More recently, similar depletion forces were found and studied
in out-of-equilibrium systems.
Dzubiella \textit{et al}~\cite{dzubiella_l_l2003} considered two
fixed big particles (or intruders) in a flowing bath of smaller particles,
all of them being modeled as soft spheres.
Their theory is based on the approximation that the perturbation
of the density field
due to the two big particles can be written as the superposition
of the perturbation due to each big particle separately.
The calculation is extended beyond this hypothesis
by~\cite{khair_b2007} for equal size colloids,
and under the assumption that interactions
between bath particles can be neglected. For simplicity, the
calculation is done when the
intruders move along their line of centers.

Related models defined on a lattice have been studied,
in which intruders undergo a biased random walk, while
the bath is made of brownian particles hopping in all
directions with equal probability.
The perturbation induced by a single intruder in a bath
of brownian particles and the resulting relation between
force and velocity distribution
was extensively studied~\cite{brummelhuis_h1989,benichou2013a}.
The case of two intruders was considered in~\cite{mejia_o2011}, in which a numerical study showed the existence
of an attractive force between the intruders resulting in
a statistical pairing.

In our case, the use of a semi-deterministic model
makes it possible to characterize {\it analytically}
the interaction between the two intruders.
We do not need to make any mean-field assumption
because we are able to consider each particular trajectory
instead of working directly with ensemble averaged wakes
(though we also predict the latter).

In their conclusion, Mejia-Monasterio and Oshanin~\cite{mejia_o2011} conjectured that the attractive interaction
that they had found between two intruders could be seen as
``an elementary act'' leading to pattern formation when
many intruders are considered.
Here we give an example of such a connection between individual and collective behaviour which can even be made explicit in the case of the alternating parallel update.

In \cite{dzubiella_l_l2003}, an experimental realization
was suggested to measure the effective depletion forces
between two big colloidal particles fixed by optical tweezers,
and placed in a flow of charged particles subject
to an electric field. Experimental studies of effective interactions were also made  with constant force driving in \cite{sokolov2011}, for particles confined on a circle.
A first step towards a physical realization of the system
studied in this paper could be to use a similar setup.
While the optical tweezers would be fixed
in the direction of the flow of the bath, there would be a servo mechanism ensuring that a constant
force perpendicular to the flow is applied on the two trapped colloids. In such a way, these
two colloids would be driven in the direction perpendicular to the electric field and localization times could \textit{a priori} be measured.

\section*{Acknowledgments}

We thank H.J. Hilhorst for inspiring discussions and for the calculation of \ref{section:rbar}.

\appendix
\section{The coefficients $\rbarfs_n$}
\label{section:rbar}

In this appendix we compute the $\rbarfs_n$ defined in section~\ref{section:ensavg} for the frozen shuffle update. 

Consider an $\symbe$ particle with phase $0$ (without loss of generality) in a flow of $\symbn$ particles, trying to hop towards site (0,0) at time 0.
This site is either empty, with probability
$1-\rhofs=\rbarfs_1$, 
or it
is occupied with probability $\rhofs dx$ by a $\symbn$
particle that we call $\symbn_1$, with phase between $1-x$ and
$(1-x)+dx$. In the second case, and using continous time,
we define $T_2$ as the interval separating the departure of
particle $\symbn_1$ from site $(0,0)$ from the arrival of its
direct follower $\symbn_2$ on the same site. From there we see
that the $\symbe$ particle will attempt to hop towards $(0,0)$
again at time $1$, whereas particle $\symbn_2$ will try at time
$1-x+T_2$. The $\symbe$ particle will therefore hop before
$\symbn_2$ if $T_2 > x$. For a fixed $x$ we get an
infinitesimal contibution to $\rbarfs_2$: $d \rbarfs_2 (x) =
\rhofs \mathrm{Prob}[T_2>x] dx$.

If the $\symbe$ particle is blocked again
by particle $\symbn_2$, we have to consider a third particle $\symbn_3$ coming in site $(0,0)$
after $\symbn_2$ after an interval $T_3$. Applying the same argument as in the previous paragraph, we see that for fixed $x$, $d \rbarfs_3 (x) = \rhofs \mathrm{Prob}[ (T_2+T_3>x) \,\mathrm{and}\, (T_2 < x)] dx$. We can finally generalize to higher numbers of blockings : $d \rbarfs_n (x) = \rhofs \mathrm{Prob}[ (T_2+\ldots+T_n>x) \,\mathrm{and}\, (T_2+\ldots+T_{n-1}<x)] dx$.

We have assumed that the incoming $\symbn$ particles
were moving in free flow with velocity $1$.
Thus, the distribution of arrival times of
the $\symbn$ particles in a given site is the
same as the distribution of injection times.
The probability density distribution of the $T_i$ is given by
\begin{equation}
\KKK(T) = a \ee^{-aT}, \qquad T > 0,
\label{defPT}
\end{equation}
where $a \geq 0$ is an inverse time which determines
the injection rate. 
The normalization reads $ \int_{T= 0}^\infty dT \KKK(T) = 1$, expressing that
every particle is necessarily followed by another particle.
One finally computes, for $n \geq 1$ 
\beq
\begin{array}{l l}
d\rbarfs_n (x) & = \rhofs \int_{T_2=0}^x \ldots \int_{T_{n-1} = 0}^{x - \sum_{i=2}^{n-2} T_i} \int_{T_n = x - \sum_{i=2}^{n-1} T_i}^\infty \prod_{i=2}^n \KKK(T_i) dT_i \\
& = \rhofs \frac{(ax)^{n-2}}{(n-2)!} \ee^{-ax}
\end{array}
\eeq

By averaging uniformly over $x$ we obtain Eq.(\ref{eq:rbarfs}).

\section{Probability distribution of the length of the queue of $\symbn$ particles.}
\label{section:queue}

We have seen in section~\ref{sec:wake_not_averaged}
that, when an $\symbe$ particle occupies a site
in column $k$ for a certain amount of time,
$\symbn$ particles accumulate below.
In this appendix we determine the probability
distribution for the length of this
queue of $\symbn$ particles in column $k$,
in the case of the frozen shuffle update. 

Let column $k$ be 
occupied by a single $\symbe$ particle during $n_k$ time steps.
We want to calculate 
$q^{n_k}_m$, 
defined as the probability that at least $m = 1,2
\ldots$ particles of type $\symbn$ have been blocked in column
$k$ by the $\symbe$ particle, before the latter leaves
the column.
We stress that this quantity is independent of the column index
$k$.

For an unperturbed flow of $\symbn$ particles,
the distribution $\KKK(T)$ defined in~\ref{section:rbar}
represents the probability of having
a time delay $T$ between the departure from one site
of a given $\symbn$ particle and the arrival of the next particle.
As there is no memory in the injection procedure,
it also represents the probability of having
a time delay $T$ between any instant where a given
site is empty and the arrival on this site of the next particle.
Using this we can write
\beq
\begin{array}{l l}
q^{n_k}_m & =  \int_{T_1 = 0}^{n_k} \int_{T_2 = 0}^{n_k-T_1} \ldots \int_{T_{m} = 0}^{n_k-\sum_{i=1}^{m-1} T_i} \prod_{i = 1}^{m} \KKK(T_i) dT_1 \ldots dT_m  \\
& = (-1)^m \sum_{l=m}^{\infty} \frac{(-an_k)^l}{l!}.
\end{array}
\eeq

Due to the relatively low density of $\symbn$ particles considered, we chose to ignore the possible relaxation of the queue occuring after the departure of the $\symbe$ particle from column $k$.
We can thus use directly the above values to compute
$R^{n_{k-1}n_k}_l(0) = \rho (1-q^{n_k}_{|l|}) + q^{n_k}_{|l|}$
for $l < 0$.
Indeed, site $(k,l)$ with $l<0$ is occupied with
probability $1$ if the queue extends beyond this site,
and with probability $\rho$ if the queue is too short
to reach this site.

\section{Rate equations for a particle in a wake}
\label{section:rateeq}

\subsection{Localization in the wake: general equations}

Consider an $\symbe$ particle in the shadow of another
$\symbe$ particle in a perpendicular flow of $\symbn$
particles, as defined in section~\ref{section:interaction}.
In the case of the frozen shuffle update, the phase of the first $\symbe$
particle creating the shadow is taken to be $0$. The phase
of the second $\symbe$ particle is denoted as $\tau_0$.
The alternating parallel update can be seen formally as a limiting case of the frozen shuffle update in which all the particles of the same type have the same phase, say $0$ for the $\symbe$ particles and $1/2$ for the $\symbn$ particles.
General equations can therefore be formulated in terms of some quantities depending on the updating scheme. These equations shall then be applied to the frozen shuffle update in~\ref{subsection:ratefs} and to the alternating parallel update in~\ref{subsection:rateap}.

We want to compute the probability $P_S(t)$ that the second particle
remains in the shadow of the first one after $t$ timesteps given by Eq.(\ref{eq:defPS}). We have shown that the shadow can be seen as a superposition of rows of two types 
$\DD$ or $\KK$ (Fig.\,\ref{fig_states}a).
In the frame of the shadow, the second $\symbe$ particle
hops from one row to the one below at each time step.

If, before hopping, the $\symbe$ particle was in state $\DW$,
then it will arrive in state $\DW$ if and only if
the target row is
of type $\DD$ (proba $1-\rho$).

If, before hopping, it was in state $\DB$,
then it will arrive in state $\DW$
only if it was blocked by a $\symbn$ particle
with phase $\tau' > \tau_0$
located just in
front
{\em and} if the target row is
of type $\DD$.
Note that the probability of the latter
is not $1-\rho$ anymore, because it is conditioned by the fact that there is a $\symbn$ particle
on the departure row, which makes it
smaller than $1-\rho$. As a result,
the probability
for this transition is
\begin{equation}
\rho \int_{\tau_0}^1  d\tau' \PQ_{\emptyset,\tau'},
\end{equation}
where $\PQ_{\emptyset,\tau'}$ denotes, for an unperturbed
vertical flow of $\symbn$ particles, the probability to have
an empty site in $(i,j)$, 
under the condition that the preceding site $(i,j+1)$ 
is occupied by a $\symbn$ particle with a phase $\tau'$.

If, before hopping, the $\symbe$ particle
was in state $\KTp$, i.e. if there was an $\symbn$ particle
of phase $\tau'$ located just in front,
then the $\symbe$ particle will arrive in state $\DW$ only if
$\tau' > \tau_0$ and if the target row is
of type $\DD$.
Again, the latter probability is conditioned by the
presence of the $\symbn$ particle.

As a result of these different contributions, we get the
first rate equation
\beq
\left.
\begin{array}{l l l}
P_W(t+1) & = (1-\rho)P_W(t) + 
\rho P_B(t)
\int_{\tau_0}^1 d\tau' \PQ_{\emptyset,\tau'} 
\\&
+ \int_{\tau_0}^1 d\tau' p_{\tau'}(t) \PQ_{\emptyset,\tau'} 
\end{array} \right.
\label{eq:rategal1}
\eeq
In this equation and the following ones, $t$ is an integer but the phases $\tau$,$\tau'$ are continuous.

Similar reasoning leads to the equations for $P_B(t)$ and $p_{\tau}(t+1)$:

\beq
\left.
\begin{array}{l l l}
P_B(t+1) & = 0 + \left[(1-\rho)\PQ_{\emptyset,\emptyset} + \rho \int_0^{\tau_0} d\tau' \PQ_{\emptyset,\tau'} \right]P_B(t)
\\&
+ \int_{0}^{\tau_0} d\tau' p_{\tau'}(t)\PQ_{\emptyset,\tau'} \\
\end{array} \right.
\label{eq:rategal2}
\eeq

\beq
\left.
\begin{array}{l l l}
p_{\tau}(t+1) & = \rho P_W(t)
\\&
+ P_B(t)
\left[ (1-\rho) \PQ_{\tau,\emptyset} \Theta ( \tau_0 - \tau)
\right.
\\&
\left.
+ \rho \int_0^\tau d\tau' \PQ_{\tau,\tau'}
\left( \Theta ( \tau' - \tau_0 ) + \Theta ( \tau_0 - \tau)
\right)
\right]
\\&
+ \Theta ( \tau_0 - \tau) \int_{0}^\tau d\tau' \PQ_{\tau,\tau'} p_{\tau'}(t)
\\&
+ \Theta ( \tau - \tau_0) \int_{\tau_0}^\tau d\tau' \PQ_{\tau,\tau'} p_{\tau'}(t),
\\
\end{array} \right.
\label{eq:rategal3}
\eeq
where $\Theta$ is the Heaviside step function. $\PQ_{\emptyset,\emptyset}$, $\PQ_{\tau,\emptyset}$ and $\PQ_{\tau,\tau'}$ are the probabilities of having a site empty / occupied by a particle with phase $\tau$ / occupied by a particle with phase $\tau$, conditioned by the fact that the site in front is empty / empty / occupied by a particle with phase $\tau'$.

We now want to calculate the probability decay
$P_S(t+1) - P_S(t)$ to stay in the wake.
One immediately sees that the coefficient of $P_W$ vanishes.
Somewhat lengthy but simple calculations
allow to simplify the remaining terms into
\begin{eqnarray}
P_S(t+1) - P_S(t) & = & -P_B(t)\left\{
(1-\rho) 
\int_{\tau_0}^1 d\tau \PQ_{\tau,\emptyset}  +
\rho \int_{\tau_0}^1 d\tau \int_{0}^{\tau_0} d\tau' \PQ_{\tau,\tau'} 
\right\}
\nonumber \\&&
- \int_{0}^{\tau_0} d\tau' p_{\tau'}(t) \int_{\tau_0}^1 d\tau \PQ_{\tau,\tau'}
\label{eq:decay}
\end{eqnarray}
One now has to replace the $\PQ$ by their explicit expressions
in the four coupled equations~(\ref{eq:rategal1}-\ref{eq:decay})
in order to solve them and evaluate the decay. This is done in~\ref{subsection:ratefs} for the frozen shuffle update and in~\ref{subsection:rateap} for the alternating parallel update.

\subsection{Frozen shuffle update}
\label{subsection:ratefs}

Using frozen shuffle update,the $\symbn$ particles are injected such that the time delay
$T$ (in continuous time) between the liberation of an entrance
site and the introduction of a new particle on this site
follows the exponential distribution $\KKK(T)$ defined in Eq.\,\ref{defPT}.

Actually $\KKK(T)$ gives the time delay distribution not only
on the entrance site, but 
also on any site if we are in the free flow phase - which is the
case considered in this subsection.
As a consequence, the transition rates $\PQ$ can
be expressed as follows:
\beq
\left\{
\begin{array}{l l l}
\PQ_{\emptyset,\emptyset} &= 1-\int_0^1 \KKK(T) dT &= \ee^{-a} = 1-\alpha \\
\PQ_{\emptyset,\tau'} &= \int_1^\infty \KKK(T-\tau') dT &= \ee^{-a(1-\tau')} \\
\PQ_{\tau,\emptyset} &= \KKK(\tau) &= a \ee^{-a\tau} \\
\PQ_{\tau,\tau'} &= \KKK(\tau-\tau') &= a \ee^{-a(\tau-\tau')} \qquad \tau > \tau' \\
\end{array} \right..
\label{eq:Qfs}
\eeq

With these expressions, the four coupled
equations~(\ref{eq:rategal1}-\ref{eq:decay})
define
completely
the time evolution for the probability $\Pfs_S(t)$
to stay in the shadow.
However, the 
integral forms for the $p_\tau$ variables make
the resolution of these equations
difficult for an arbitrary density.
Equations become much simpler in 
the limit of small densities, for which one has $p_\tau =
O(\rhofs)$ and $\alpha,a = \rhofs + O(\rhofs)$.
Indeed, in this case, it is possible to get rid of
integrals by introducing new variables $\Pfs_{< \tau_0} (t) \equiv \int_{\tau = 0}^{\tau_0} p_\tau (t) d\tau$ and $\Pfs_{> \tau_0} (t) \equiv \int_{\tau = \tau_0}^{1} p_\tau (t) d\tau$.
Then Eqs.(\ref{eq:rategal1}-\ref{eq:decay}) become
Eqs.(\ref{eq:ratelowd}-\ref{eq:pfs_lowd}) after some 
calculations.

\subsection{Alternating parallel update}
\label{subsection:rateap}

Here we give the expressions of the $\PQ$ for alternating parallel update.
In contrast with the frozen shuffle update for which several $\symbn$ particles could occupy successive sites, here two $\symbn$ particles are separated by at least one hole in the direction
of propagation.
The conditional probabilities therefore read
\beq
\left\{
\begin{array}{l l l}
\PQ_{\emptyset,\emptyset} &=& 1-\alpha \\
\PQ_{\emptyset,\tau'} &=& \delta(\tau'-1/2) \\
\PQ_{\tau,\emptyset} &=& \alpha \delta(\tau-1/2) \\
\PQ_{\tau,\tau'} &=& 0 \\
\end{array} \right.,
\label{eq:Qap}
\eeq
where $\delta(x)$ is Dirac's delta distribution. Injecting in Eqs.(\ref{eq:rategal1}-\ref{eq:decay}) and using $\rhoap = \alpha/(1+\alpha)$ finally gives Eqs.(\ref{eq:ratealtpar}), which are true for all densities.

\end{document}